\documentclass{aa}
\usepackage{graphicx}

\begin{document}

\title{Variability of the H$\alpha$ emission of Cygnus~X-1 and its connection with the soft X-ray radiation}

\author{A.E.Tarasov\inst{1}
  \and C.Brocksopp\inst{2}
  \and V.M.Lyuty\inst{3}}

\institute{Crimean Astrophysical Observatory and Isaac Newton Institute of Chile, Crimean branch, Nauchny, Crimea, 98409 Ukraine.
  \and Astrophysics Research Institute, Liverpool John Moores University, Twelve Quays House, Egerton Wharf, Birkenhead CH41 1LD, UK
  \and Sternberg Astronomical Institute, Universitetskii pr.13, Moscow, 119899 Russia.}

\mail{tarasov@crao.crimea.ua}

\date{Received date: 09-07-02 / Accepted date: 28-01-03}

\titlerunning{Variability of the H$\alpha$ emission of Cygnus X-1}

\abstract{
High-resolution H$\alpha$ monitoring of Cyg X-1, HD226868 was carried out during 1996--2002 and the resultant spectra analysed in conjunction with 1.5--12 keV X-ray monitoring. We demonstrate that the H$\alpha$ line-profiles have complex variability on different timescales, controlled in particular by the orbital period and the focused wind model of mass loss. We find that long-term variability of the mass loss by the supergiant and short-term variability due to clumpy structure of the stellar wind dominate during the low/hard X-ray state and that X-ray photoionization has a relatively small influence on the line-profile shape and EW variability. During the high/soft X-ray state and flaring the effect of photoionization the line-profile and EW of H$\alpha$ increases but is still unable to describe the loose anti-correlation between EW and the low energy X-ray emission. We propose that variability of the mass loss by the supergiant can change wind velocities in the Str\"omgren zone around the accretion disc of the secondary, leading to an increase in accretion rate through the disc.
}

\maketitle

\section{Introduction}

\begin{table*}
\caption{Equivalent widths and orbital phase values of the H$\alpha$ line-profiles. Spectra marked by an asterisk have been plotted in Figs. 1 and 2.}
\begin{center}
\begin{tabular}{llllllllllll}
\hline
MJDh&Phase& EW & MJDh &Phase& EW& MJDh &Phase& EW& MJDh &Phase& EW\\
$-50000$&&(\AA)&$-50000$&&(\AA)&$-50000$&&(\AA)&$-50000$&&(\AA)\\
\hline
273.4990  & 0.830 &  0.234 & 873.6349  & 0.000 & -1.488 & 1394.4955* & 0.014 & -0.910 &  2071.3417  & 0.883 & -0.433\\    
274.4101* & 0.993 &  0.220 & 941.5449* & 0.128 & -2.315 & 1395.3098* & 0.160 & -1.044 &	 2091.4221  & 0.469 & -0.364\\    
275.3955* & 0.169 &  0.371 & 961.6570* & 0.719 & -1.092 & 1395.5007  & 0.194 & -1.013 &	 2092.4150  & 0.646 & -0.290\\    
325.4365* & 0.105 & -0.332 & 992.4580  & 0.220 & -0.615 & 1396.2702  & 0.331 & -0.771 &	 2095.3404  & 0.169 & -0.484\\    
326.4219* & 0.281 &  0.057 & 993.4146* & 0.390 & -0.679 & 1396.4955  & 0.371 & -0.891 &	 2096.3794  & 0.354 & -0.580 \\   
401.1547  & 0.626 & -0.530 & 1003.4171  & 0.177 & -0.944 & 1397.2914  & 0.513 & -0.525 & 2097.4344  & 0.543 & -0.270 \\   
606.5311* & 0.302 & -0.594 & 1004.4587  & 0.363 & -0.845 & 1397.5278* & 0.556 & -0.362 & 2149.4207  & 0.827 & -0.402 \\   
607.5300  & 0.480 & -0.333 & 1006.4288  & 0.714 & -0.870 & 1398.3053* & 0.694 & -0.432 & 2164.3105* & 0.485 & -0.247\\    
612.4194* & 0.353 & -0.253 & 1007.4263* & 0.893 & -0.866 & 1398.5396  & 0.736 & -0.211 & 2172.3634* & 0.923 & -0.024\\    
615.4088  & 0.887 & -0.824 & 1008.4263* & 0.071 & -0.836 & 1420.4177  & 0.643 & -0.390 & 2177.2954  & 0.804 & -0.131\\    
623.4019  & 0.315 & -0.612 & 1009.4183* & 0.248 & -0.781 & 1421.4212* & 0.822 &  0.088 & 2186.3181* & 0.415 &  0.025\\    
624.4133* & 0.495 & -1.049 & 1010.4148* & 0.426 & -1.013 & 1422.4295  & 0.001 & -0.084 & 2187.2748  & 0.586 & -0.145\\    
625.4060  & 0.673 & -0.634 & 1015.4179  & 0.320 & -0.530 & 1443.2268* & 0.716 &  0.196 & 2188.2104* & 0.753 & -0.145\\    
626.3952  & 0.849 & -0.431 & 1016.4356* & 0.501 & -0.480 & 1444.2386* & 0.897 &  0.058 & 2211.1366  & 0.847 & -0.269\\    
633.5005* & 0.118 & -1.082 & 1020.4175  & 0.212 & -0.581 & 1455.3857  & 0.888 &  0.336 & 2328.5642  & 0.817 & -0.327 \\   
634.4984* & 0.296 & -0.790 & 1021.4126  & 0.390 & -0.294 & 1475.2194* & 0.430 & -0.139 & 2365.4850  & 0.410 & -0.298 \\   
635.5022  & 0.475 & -0.502 & 1026.2785* & 0.259 & -0.662 & 1476.2507  & 0.614 & -0.003 & 2369.5055  & 0.128 & -0.056 \\   
647.3683* & 0.594 & -1.067 & 1123.3736  & 0.598 & -1.176 & 1482.1972* & 0.676 & -0.332 & 2394.4212* & 0.578 & -0.159 \\   
648.4057* & 0.780 & -1.022 & 1238.6054  & 0.176 & -0.932 & 1766.3050  & 0.411 &  0.052 & 2398.4434  & 0.296 & -0.384 \\   
649.4439  & 0.965 & -0.861 & 1264.6038  & 0.818 & -0.536 & 1769.3152  & 0.948 & -0.866 & 2401.4410  & 0.831 & -0.148 \\   
650.4508  & 0.145 & -1.277 & 1293.4793* & 0.975 & -0.194 & 1770.2515  & 0.115 & -1.014 & 2409.4680  & 0.265 & -0.374\\    
651.4460* & 0.323 & -0.999 & 1294.4765  & 0.153 & -0.698 & 1771.3322* & 0.308 & -0.290 & 2410.4010  & 0.431 & -0.245\\    
652.3862  & 0.490 & -0.607 & 1305.4519  & 0.113 & -0.444 & 1858.1293* & 0.808 &  0.215 & 2411.4400* & 0.616 & -0.387\\    
653.4612  & 0.682 & -0.448 & 1308.4365* & 0.646 & -0.034 & 1863.1569* & 0.706 & -0.031 & 2419.3912* & 0.037 & -0.021\\    
658.4077* & 0.566 & -0.149 & 1328.4503* & 0.220 & -0.523 & 1864.1541* & 0.884 &  0.569 & 2446.3902  & 0.858 & -0.083\\    
660.3157  & 0.907 & -0.170 & 1332.4289* & 0.930 & -0.982 & 1865.1716* & 0.066 &  0.500 & 2447.3374* & 0.027 &  0.355\\    
661.3577* & 0.093 & -0.303 & 1333.4077  & 0.105 & -0.873 & 1866.1684  & 0.244 &  0.130 & 2452.3074  & 0.914 &  0.270\\    
663.4070* & 0.459 & -0.228 & 1336.4273  & 0.645 & -0.428 & 1867.1610* & 0.421 & -0.160 & 2456.3341  & 0.634 & -0.278\\    
666.3997  & 0.993 & -0.496 & 1381.4422  & 0.683 & -0.542 & 1868.1514  & 0.498 & -0.492 & 2481.4483  & 0.119 & -0.271\\    
667.4112* & 0.174 & -0.244 & 1382.4352  & 0.860 & -0.594 & 1887.1860  & 0.997 &  0.061 & 2487.4750  & 0.195 & -0.253\\    
668.3608  & 0.343 & -0.371 & 1392.3289* & 0.627 & -0.781 & 1986.5603  & 0.743 & -0.858 & 2488.4117  & 0.362 &  0.072\\    
669.3799  & 0.325 & -0.392 & 1392.5227* & 0.662 & -0.711 & 2015.5363* & 0.918 & -0.618 & 2510.3777  & 0.285 &  0.179\\    
675.2605  & 0.575 & -0.296 & 1393.3185* & 0.804 & -0.775 & 2063.4307  & 0.471 & -0.710 & 2515.3919  & 0.180 & -0.377\\    
676.4133* & 0.781 & -0.644 & 1393.5074  & 0.838 & -0.786 & 2069.3228  & 0.523 & -0.202 & 2517.4513  & 0.548 & -0.105\\    
677.4112  & 0.959 & -0.678 & 1394.2875* & 0.977 & -0.896 & 2070.3660  & 0.709 & -0.749 & 2534.3777  & 0.581 &  0.203\\    
764.2573  & 0.468 & -0.408 &            &       &        &            &       &        & 2569.2194  & 0.793 & -0.646\\
\hline         
\end{tabular}  
\end{center}   
\end{table*}

Cyg X-1 has been one of the most well-studied X-ray sources since its discovery in 1965 (Bowyer et al. 1965) and identification with the O9.7Iab supergiant HD226868\, (Bolton 1972; Webster \& Murdin 1972). It provided the first observational evidence for the existence of black holes when it was found to be a spectroscopic binary with an orbital period of 5.6 days. The ephemeris was revised a number of times and a recent estimation based on spectroscopic and photometric orbital variability is presented by Brocksopp et al. (1999a). The masses of the two components were calculated as 17.5\,$M_{\odot}$ for the supergiant and 10.1\,$M_{\odot}$ for the black hole, with adopted system inclination $i$=35\degr (Herrero et al. 1995).

Cyg X-1 is a detached double system but the optical component is very close to filling its Roche lobe. Typically of high mass stars, HD22686 is thought to emit a stellar wind, which appears to be attracted towards the black hole as in the focused wind model of Friend \& Castor (1982; Gies \& Bolton 1986; Sowers et al. 1998). The optical spectrum of the system has only two well-pronounced emission lines: H$\alpha$ and He\,{\sc ii}\,$\lambda$4686, but high-resolution spectroscopy detects line-asymmetry in most of the hydrogen and red He\,{\sc i} lines (Canalizo et al. 1995) which thus cannot be used for precise determination of the orbital elements. Gies \& Bolton (1986) and Ninkov et al. (1987) found that the He\,{\sc ii}\,$\lambda$4686 emission component has a radial velocity curve that is best interpreted as emission originating between the stars. The H$\alpha$ profile appears to have at least two components; P Cygni emission moving with the orbital motion of the primary star and a second emisison component which follows the quite different radial velocity curve of the He\,{\sc ii}\,$\lambda$4686 emission line and which is formed in the focused wind flow between the stars (Ninkov et al. 1987; Sowers et al. 1998).

Cyg X-1 demonstrates sporadic flaring at soft X-ray energies (2-12 keV) on different timescales, from days to months and years (Brocksopp et al. 1999b). It is believed that the source of that variability is located near the black hole companion and is connected with the accretion disc. A supergiant wind is also variable on different timescales from days to years (Kaufer et al. 1996; Rivinius et al. 1997) and the variability of the emission lines and the soft X-rays should be somehow connected because they both depend on mass transfer via the wind. Discovery of any correlation between the stellar wind and the X-ray variability is important for understanding the physical processes of mass transfer through the disc and onto the black hole.

Long-term monitoring of variability of the stellar wind from the supergiant of Cyg X-1 has not been previously attempted. Nonetheless some orbit-independent variability of the equivalent widths of the He\,{\sc ii}\,$\lambda$4686 emission line was found by Ninkov et al. (1987) and of the emission component of H$\alpha$ by Voloshina et al. (1997). The latter also discovered a decrease of EW accompanying the period of high/soft state behaviour in 1996 (see also Brocksopp et al. 1999b). Thus we have monitored the H$\alpha$ line at the Crimean Astrophysical Observatory (CrAO) since the 1996 X-ray spectral state change, when a strong increase in soft X-rays was found by the Rossi X-ray Timing Explorer ({\it RXTE}); our observations were supported by long-term photometric monitoring provided by Sternberg Astronomical Institute of Moscow.

\begin{figure*}
\centering
 \includegraphics[width=13cm]{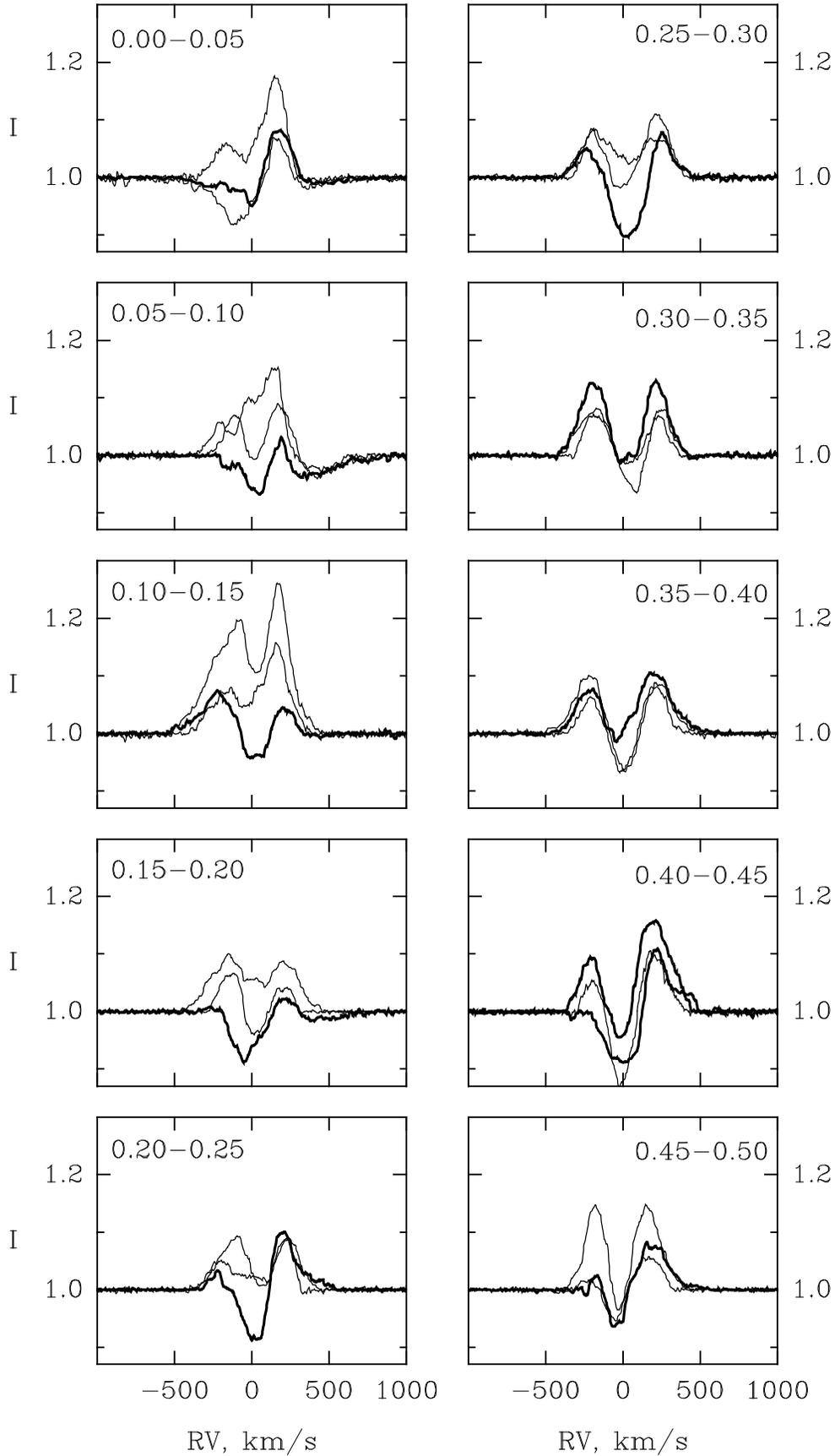}
\caption{Selected H$\alpha$ profiles for orbital phases 0.00-0.50, determined according to the ephemeris of Brocksopp et al. (1999). Each box contains three sample spectra, chosen at random in terms of continuum or X-ray activity but so as to include extremely strong, weak and intermediate intensity examples of the H$\alpha$ emission line in each case (plotted spectra are indicated in Table 1). Spectra drawn with a thick line are those obtained during periods of X-ray activity.}
\label{h3823F1.eps}
\end{figure*}

\begin{figure*}
\centering
 \includegraphics[width=13cm]{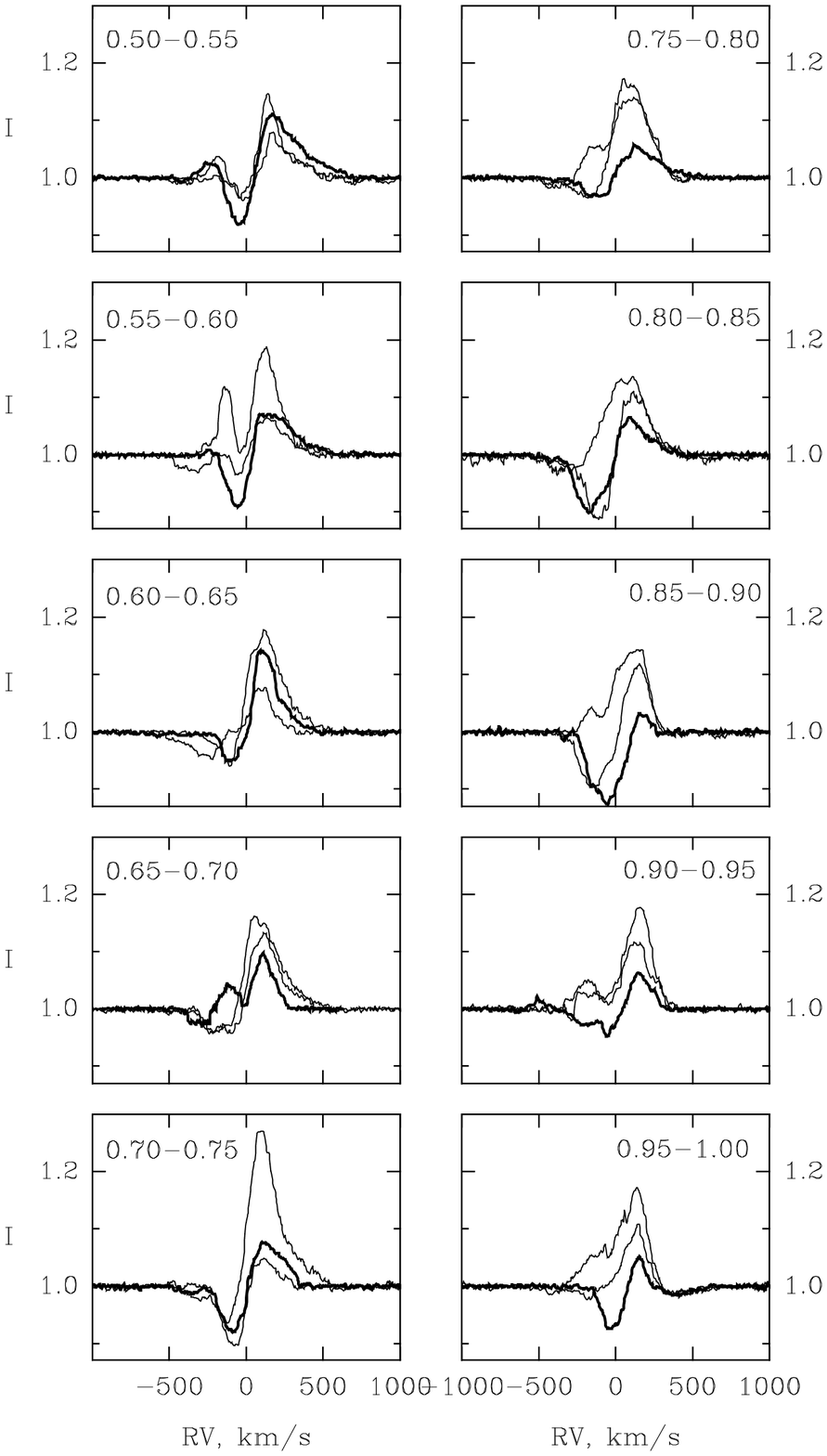}
\caption{Selected H$\alpha$ profiles for orbital phases 0.50-1.00, chosen as in Fig. 1}
\label{h3823F2.eps}
\end{figure*}

\section{Spectroscopic observations and data reduction}

The spectroscopic monitoring of Cyg~X-1 took place over seven years from 1996 to 2002, during which time we obtained a total of 142 spectra. We used the Coud\'e focus of the 2.6-m telescope of the CrAO and the detector was an Electronix CCD array (1024$\times$260 pixels). All observations were performed in the second order of a diffraction grating with a reciprocal dispersion 3\,\AA/mm. The spectral resolution was about 35000 and during each observation we obtained a 60\,\AA\,  spectrum. The exposure time ranged from 30 to 45 minutes, depending on weather conditions resulting in a signal-to-noise ratio of 100 or better. In most cases two or three spectra were obtained during each night and then added for improved signal-to-noise and cosmic ray subtraction.

The spectra were reduced using standard techniques for dark-current subtraction and flat-field division. The subsequent reduction included removal of spurious spikes produced by cosmic-rays, sky-background subtraction, water vapor line subtraction using non-emission fast-rotating early type stars and filtering of the spectra by the moving-average method using three points. For the wavelength calibration we used Th\,Ar comparison spectra obtained immediately before and after the target observation. The accuracy of the calibration was 1.5 km/s or better. All spectra were corrected to the baricentre of the solar system, both in wavelength and time of mid-exposure. The continuum normalization was provided by fitting of a fifth degree polynomial. 

As a final step, equivalent widths (EW) of each spectrum were measured. Estimated errors of EW measurements were better than 0.1\AA. The results are presented in Table 1.

During the period of our observations the system was monitored by {\it RXTE}. In order to compare variability of H$\alpha$ with the X-ray variability, we use
 the All Sky Monitor (ASM) public archive data from the web. The ASM observations were made in the three energy bands:  A (1.5--3 keV), B (3--5 keV) and C (5--12 keV). A detailed description of the ASM, including calibration and reduction, is published in Levine et al. (1996). The daily mean data were converted from counts to energy units ($\mbox{keV}\,\mbox{cm}^{-2}\mbox{s}^{-1}$) using transformation formulae obtained from Zdziarski et al. (2002).

\section{H$\alpha$ line-profile variability}

Previous study of the emission spectrum of Cyg X-1 shows that the H$\alpha$ line-profile has strong variability on the 5.6-day orbital period (Gies \& Bolton, 1986; Ninkov et al. 1987; Sowers et al. 1998). However our observations show high variability of the emission line intensity that is phase-independent and which was not noticed in the previous observations (which took place over a shorter time range). Since mass loss by the optical component is the only source of accretion, any instability of the stellar wind will lead to a variable mass accretion though the response of the accretion disc and that will have some impact on the X-ray emission.

Such complex variability makes analysis of the line-profile variations much more difficult to interpret because the shape and strength of the H$\alpha$ emission profile depends on a number of factors:
\begin{enumerate}
\item[i)] phase of the orbital period (wind is noticeably collimated in the direction of the secondary)
\item[ii)] variability of the stellar wind density due to variable mass loss rate by the supergiant
\item[iii)] clumpy structure of the stellar wind providing discrete absorption components (DAC)
\item [iv)] time-variable and phase-locked photoionization of the stellar wind by the soft X-ray radiation from the accretion disc of the secondary component
\end{enumerate}

In order to overcome this we have divided our spectra into orbital phase groups spanning 0.05 in phase and analysed each group separately according to processes that may have influenced them. Figs. 1 and 2 show three sample line-profiles for each group, chosen at random in terms of continuum or X-ray activity but so as to include extremely strong, weak and intermediate intensity examples of the H$\alpha$ emission line in each case (plotted spectra are indicated in Table 1). Spectra drawn with a thick line are those obtained during periods of X-ray activity (i.e. intensity of the 1.5--3 keV X-rays in the exceed 4 $\mbox{keV}\,\mbox{cm}^{-2}\mbox{s}^{-1}$) and it appears that the X-ray variabilty has little effect on the profiles (see Sect. 4).  

\begin{figure*}
\centering
 \includegraphics[width=16cm]{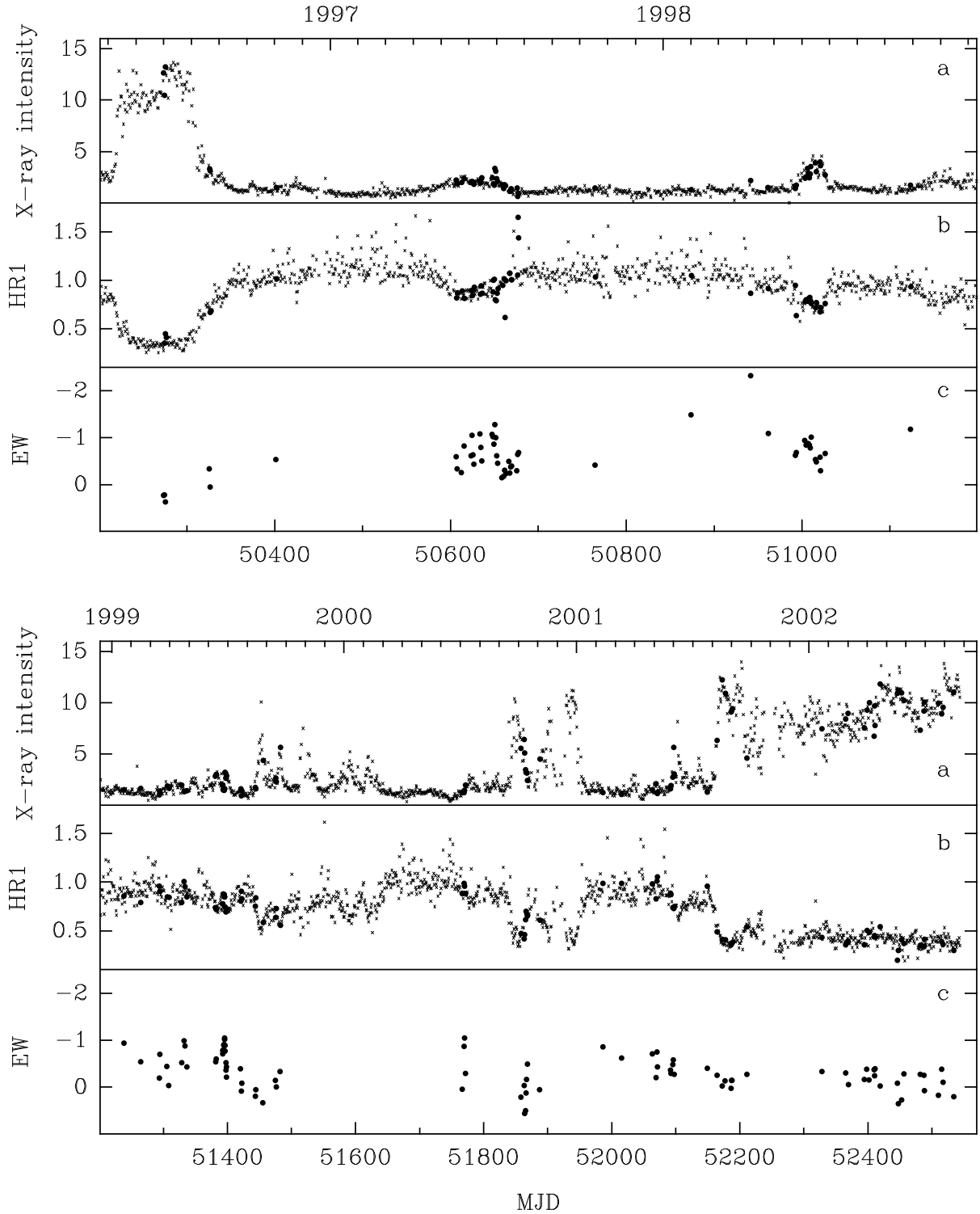}
\caption{Long-term variability of: a)X-ray ASM intensity in the 1.5--3 keV band ($\mbox{keV}\,\mbox{cm}^{-2}\mbox{s}^{-1}$); b) hardness ratio (HR1) of the soft X-ray radiation, (3--5 keV)/(1.5--3 keV); c) EW of H$\alpha$ (\AA). Filled circles on Figs. 3a \& 3b indicated X-ray measurements obtained simultaneous with H$\alpha$ observations.}
\label{h3823F3}
\end{figure*}

Figs. 1 and 2 show that the orbital variability of the line profile is mostly in agreement with the focused wind model; there are two emission components as demonstrated by Sowers et al. (1998). One forms a relatively stable P\,Cyg structure, which is produced by the stellar wind from the hot supergiant and moves around the spectrum in the phase of the orbital period. The other emission component originates between the stars in a focused wind flow from the supergiant to the unseen companion. The orbital movement of this component of the line-profile is different since it appears in the collimated stellar wind close to the L1 point, such that it is blue-shifted during the phase range 0.10--0.60 and merged with the P\,Cyg emission component at other times.

\begin{figure}
\center
 \includegraphics[width=6cm]{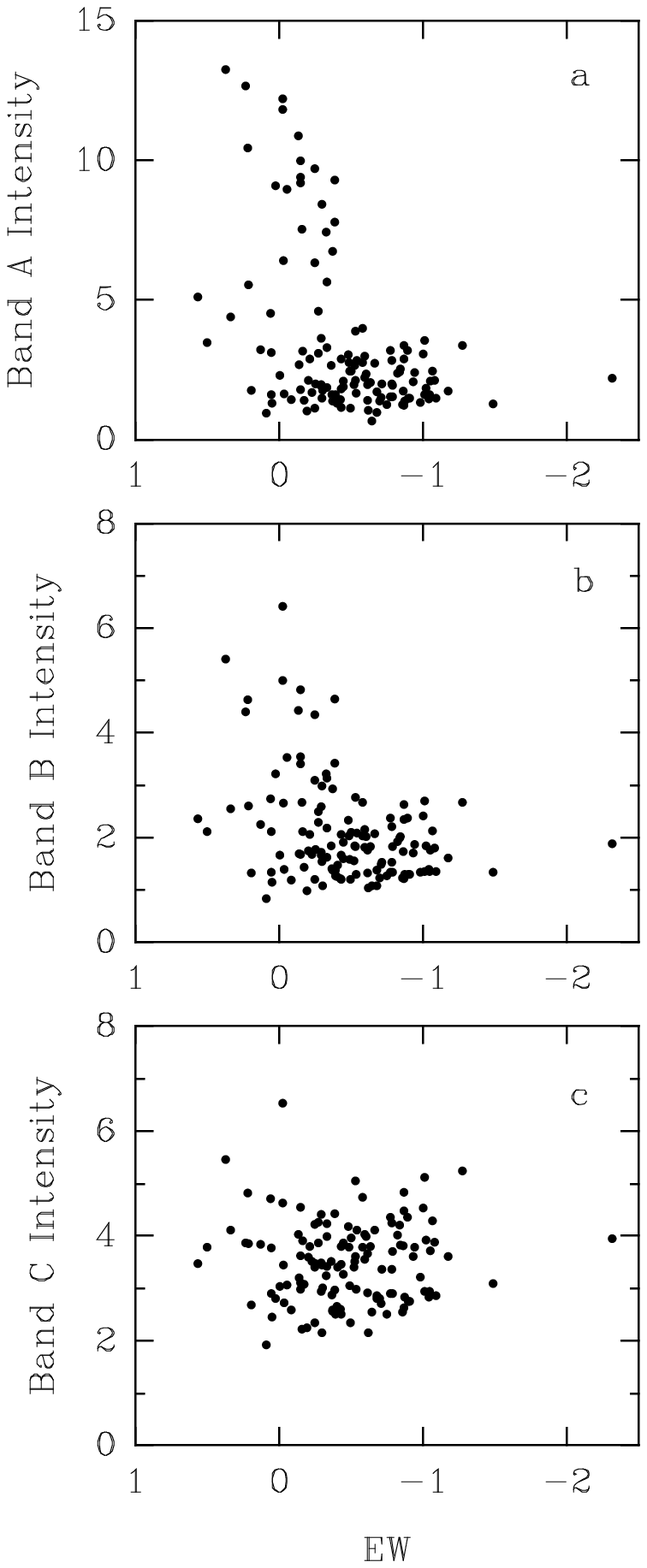}
\caption{Soft X-ray intensities plotted against the EW of H$\alpha$ for each of the three X-ray energy bands. X-ray intensities. Characteristic behaviour can be seen for some (but not all) ranges of EW for Band A and not at all for Band C.}
\label{h3823F4.eps}
\end{figure}

\begin{figure}
\center
 \includegraphics[width=6cm]{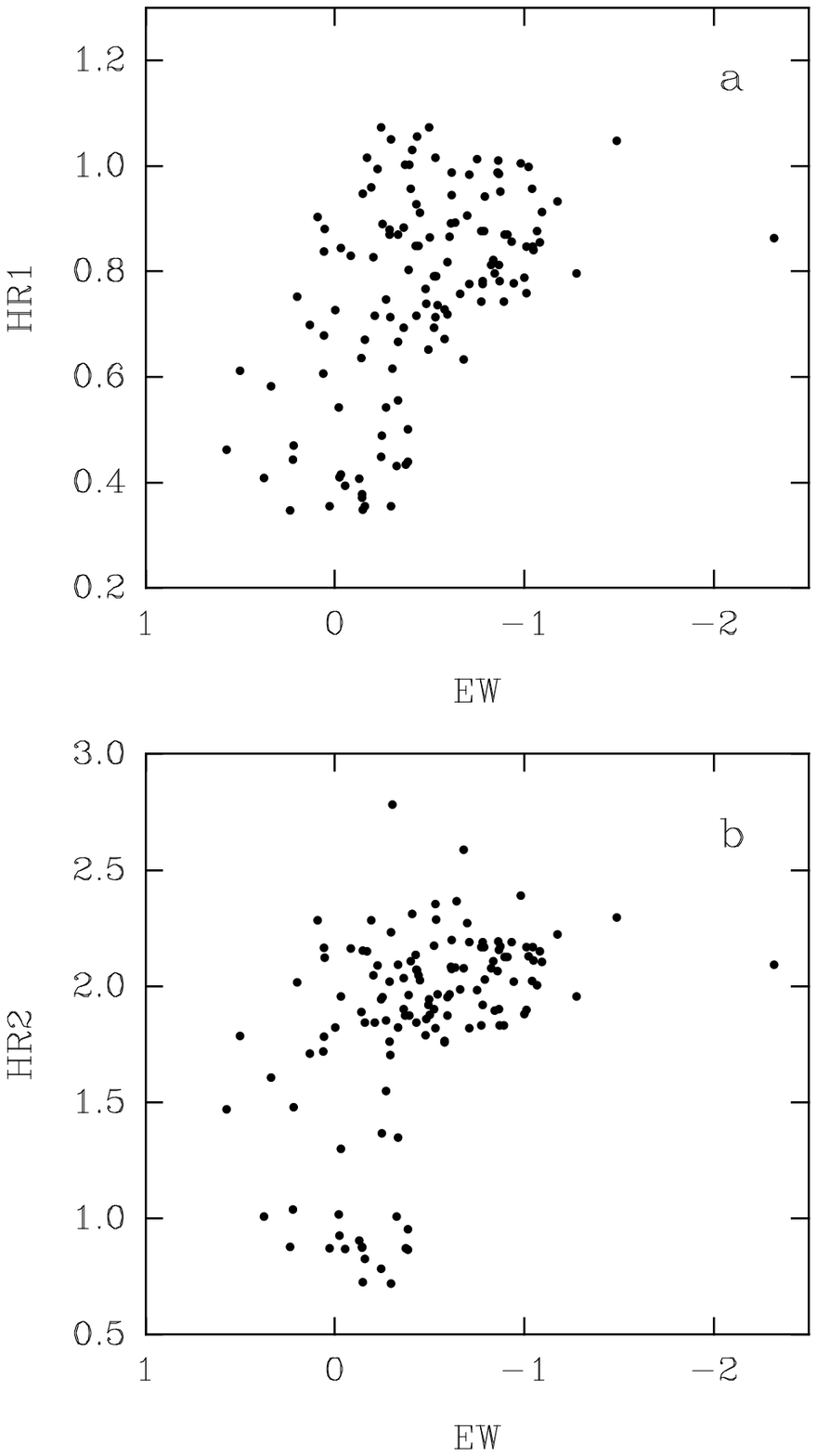}
\caption{Relationship between the hardness of the low-energy X-ray spectrum and EW of H$\alpha$; a) HR1 -- (3--5 keV)/(1.5--3 keV) and b) HR2 -- (5--12 keV)/(3--5 keV). This apparent correlation is clearly more significant for HR1 (top) than HR2 (bottom).}
\label{h3823F5}
\end{figure}

Irregular variability of the line-profiles is clearly seen during all phase intervals; sometimes (phases: 0.25-0.35, 0.50-0.55, 0.70-0.75) there is an increase or decrease in the intensity of the whole line but for the majority of phase intervals irregular variation of the blue part of the profile seems to dominate. These variations can be connected with the variability of either the collimated wind component or with the absorption part of the P\,Cyg line-profile from the stellar wind of the supergiant. However since both emission components reflect variable mass loss from the optical star, it is most probable that they are connected with one another.

Fig. 3 shows long-term light curves for the $RXTE$ intensity (Band A), hardness ratio (HR1 = Bands B/A) and H$\alpha$ EW. As with the line profiles it is clear that there is complex variability in the EW measurements, on both short ($\sim$days e.g. MJD 51392--51398) and long ($\sim$few weeks e.g. MJD 51980--52200) timescales but the sampling is often insufficient to be conclusive. However, it is certain that this variability is not connected with the orbital period and instead seems to reflect some other mechanism(s) within the system.

\section{Connection with the soft X-ray variability}

The ASM data plotted in Fig. 3 demonstrate that after the high/soft state period (i.e. the major outburst in the softer X-ray bands) in 1996 (MJD 50100--50400) the subsequent flares increased in both intensity and softness from year to year, reaching a maximum in 2002 with another high/soft state period. These flares have a semi-regular nature with characteristic occurrence time of $\sim$400 days. 

The variability of the EW and ASM X-ray flux are often loosely correlated. For example the most active X-ray periods MJD 50200--50350 and 52100--52500 were both accompanied by a decrease in EW; similarly the more short-lived flares during MJD 51850--51900 and 51440--51480. However a decrease in EW is not proportional to the amplitude of X-ray flares and is not always associated with episodes of X-ray activity. Furthermore the EW of H$\alpha$ decreased monotonically from $-1.1$\,\AA\, on MJD 51397 to 0.3\,\AA\, on MJD 51456 (Fig. 3b) {\em prior} to the X-ray flare, reaching a minimum simultaneously with the small X-ray peak.

To understand the situation better, we combined the EW of H$\alpha$ with the three different ASM spectral bands. Fig. 4 shows that the relationship between the EW and the softer X-ray emission depends strongly on the nature of the spectral state; X-ray flares appear only if the EW drops below $-0.4$\,\AA. The softest X-ray spectral band (A) is most sensitive to the H$\alpha$ variability such that a decrease in EW will {\em probably}, although not necessarily, be associated with an increase in X-ray flux as seen in the lightcurves; on the contrary the hardest X-ray band (C) appears insensitive to the EW.

\begin{figure*}
\center
 \includegraphics[width=12cm]{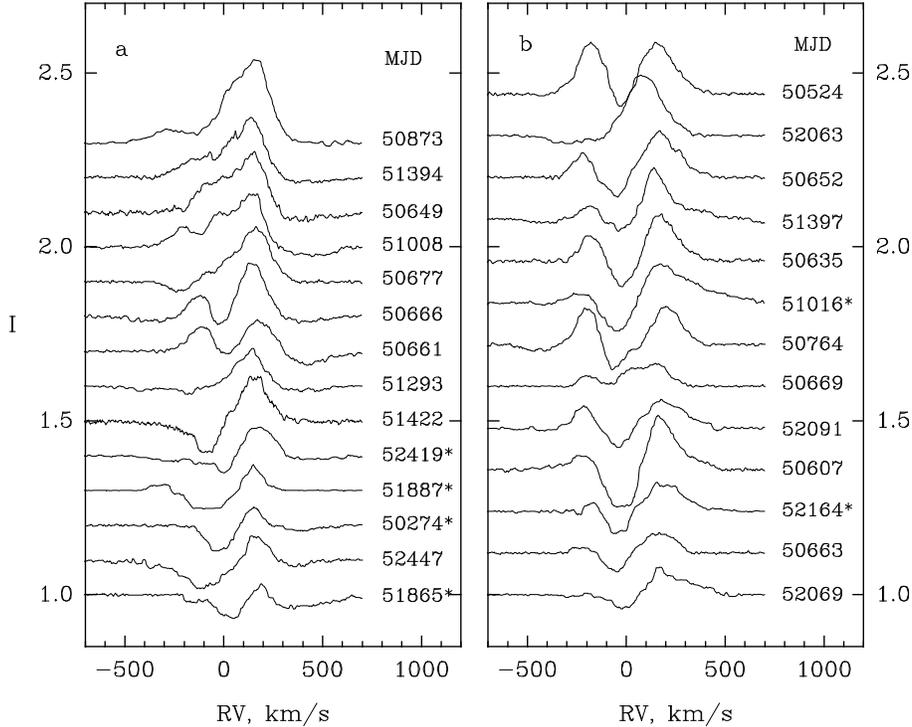}
\caption{All our H$\alpha$ line profiles obtained during phase intervals 0.95-1.10 (a) and 0.45-0.55 (b). Profiles marked by an asterisk were obtained during the time of X-ray activity, when intensity in the band A (1.5-3.0 keV) were more than 4 $\mbox{keV}\,\mbox{cm}^{-2}\mbox{s}^{-1}$.}
\label{h3823F6}
\end{figure*}

The hardness of the X-ray spectrum can be a useful parameter for understanding the relation between X-ray emission and emission in the H$\alpha$ line -- the EW and X-ray hardness ratio appear to be correlated for both X-ray colours (Fig. 5), although that of HR1 (i.e. the softer colour) is more significant. The scattering of the data in Fig. 5 is significantly larger than the estimated errors of the observations, both in EW and X-ray and thus some additional factors must play an important role in the connection between optical and X-ray variability.

Comparing Figs. 4 and 5 it is possible to conclude that the loose correlation between the H$\alpha$ and X-ray emission is more significant for the X-ray spectral hardness than for the intensity. 

\section{Discussion}

A number of sources could be responsible for the presence of some relationship between the variability of the H$\alpha$ EW and low-energy X-ray radiation; these include  (i) photoionization of the neutral gas by the soft X-ray radiation, (ii) X-ray flares and state transitions of the accretion disc due to variable mass transfer via the stellar wind and/or (iii) variability of X-ray radiation due to variable column density of neutral gas in the line of sight. 

\subsection{Photoionization vs. mass-loss variability}

Unlike single hot supergiant stars, Cyg X-1 has an additional source of ionisation in the form of X-ray radiation from the accretion disc of the secondary. However, due to orbital motion, it is phase locked with the orbital period and can be identified relatively easily. This is illustrated in Fig. 6 where we display all spectra in the orbital phase intervals 0.95--1.10 (left) and 0.45-0.55 (right). During the first of these phase intervals the black hole is situated behind the optical component, with the red component of the H$\alpha$ profile formed in the part of the envelope which is directed towards the X-ray source (resulting in a superposition of the envelope and focussed wind emission components). During the second phase interval the star is behind the black hole and the red emission is no longer under the influence of the X-ray source.

Those spectra obtained during periods of X-ray activity have been indicated in Fig. 6 by an asterisk and show clearly that the effects of photoionisation by the X-ray emission do not dominate the shape of the emission line. However, comparison of the red wing of the red emission peak at different orbital phases shows that photoionisation does have some influence on the line-profile. At phase 0.0 the red wing has negligible extension and the photospheric line profile is often seen; this can be attributed to photoionization by the X-rays since the red wing of the emission forms bewteen the star and the black hole. At phase 0.5 the red wing forms on the far side of the star where it is unaffected by the X-ray radiation and can be extended up to 700 km/s. During both phase intervals plotted in Fig. 6 the aperiodic part of the wind variability has a strong influence over the shape of the H$\alpha$ line. Often this sporadic inhomogeneity of the wind or long-term variability of mass loss is so high that it prevails over the (focused wind model controlled) orbital line-profile variability. This can occur at any phase interval of the orbital period and independently of X-ray activity.

In the top panel of Fig. 7 we plot the EW values against orbital phase, using filled circles for the data obtained during periods of X-ray activity (intensities in the 1.5--3 keV band exceed 4 $\mbox{keV}\,\mbox{cm}^{-2}\mbox{s}^{-1}$). While the quiescent (and combined) data show only scatter, the scatter for the filled circles (i.e. epochs of X-ray flaring) is much more constrained.

\begin{figure}
\center
 \includegraphics[width=8cm]{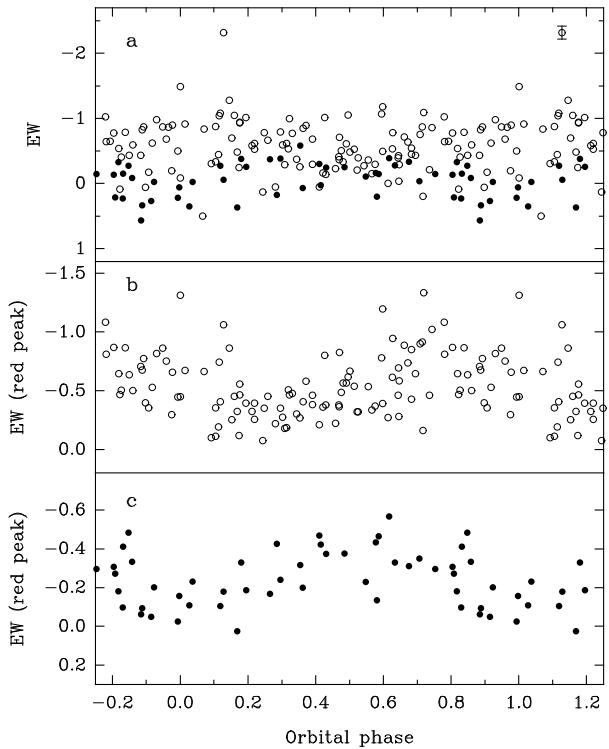}
\caption{(a) Variability of the equivalent width of the H$\alpha$ line with the phase of orbital period. Open circles -- EWs obtained outside X-ray activity; filled circles - EWs obtained during X-ray activity, (intensities of the soft X-ray in the 1.5-3 keV band exceed 4 $\mbox{keV}\,\mbox{cm}^{-2}\mbox{s}^{-1}$). Typical error of EW measurements are shown for one of points. (b) Similar plot showing the EW of the red emission peak only, during non-flaring episodes. (c) Similar plot showing the EW of the red emission peak only, during flaring episodes.}
\label{h3823F7}
\end{figure}

The presence of the orbital variability is best seen in the EW of the red emission peak alone. The highly scattered wave in Fig. 7b (non-flaring episodes) is in full agreement with the focused wind model and increasing of EW at phase 0.6--0.9 can be seen as the collimated wind component merges with the P Cyg wind component (see Fig. 2). The photoionization shown in Fig. 6 is relatively small in amplitude and the large scattering of the points in Fig. 7b can be best-described by the short- and long-term variability of the stellar wind. Fig. 7c (flaring episodes) shows that photoionisation dominates during phases 0.6--1.3 when the X-ray source is located close to where the red emisison peak is forming. Both curves on Figs. 7b \& 7c are statistycally significant -- the full amplitudes of EW variability are exceed more than ten times standard deviations of the data. This is in agreement with the orbital variability of the resonance UV lines studied by Theves et al. (1980) and van Loon et al. (2001). The resonance UV lines form in a more extended region (which includes the  Roche lobe of the secondary) than H$\alpha$ and are significantly influenced by photoionization even during quiescence.

\begin{figure}
\center
 \includegraphics[width=6cm]{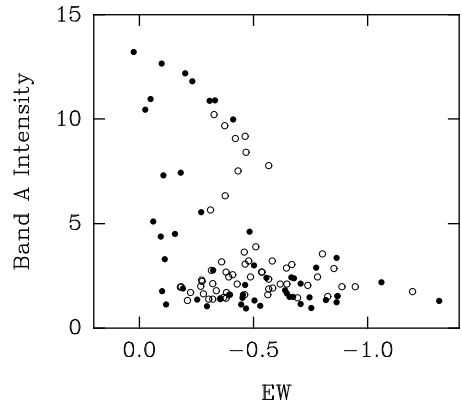}
\caption{Soft X-ray intensities (1.5-3 keV energy band) plotted against the EW of H$\alpha$ red emission peak. Filled circles -- phase interval of the orbital period $0.0\pm0.2$; open circles phase interval $0.5\pm0.2$.}
\label{h3823F8}
\end{figure}

The complex relationship between the X-rays and H$\alpha$ EW that was seen in Figs. 4 and 5 cannot be explained purely in terms of time-variable photoionisation. Plotting EWs of the red emission peak against the 1.5--3 keV X-rays for phase intervals $0.0\pm0.2$ (filled circles) and $0.5\pm0.2$ (open circles) shows a similar plot to Fig. 4 for each phase interval (Fig. 8). If photoionization dominated then we would expect to see correlated behaviour in each phase interval; instead a large scatter and independent variations are still present.

The decay of EW {\em prior} to the X-ray flare of MJD 51456 would suggest that (at least in some instances) a decrease in mass loss can trigger an X-ray flare. A reduction in mass-loss would increase the size of the Str\"omgren (i.e. photoionisation) zone surrounding the X-ray source (van Loon et al. 2001). This would increase the accretion rate through the disc and result in X-ray activity, thus increasing the size of the Str\"omgren zone further. In the case of the 2001--2002 high/soft state of Cyg X-1, once the EW return to $-1.5$ -- $-2.0$ \AA~ the Str\"omgren zone would be compressed to its normal size, as was the case after the 1996 high/soft state . However, this would appear not to be the case for all flares and optical observations with improved time resolution are needed to investigate this further.

\subsection{Variability of the column density toward secondary}
  
It is well known that the winds of early type stars are driven by resonance line scattering of radiation in the stellar envelope. This radiative driving mechanism is inherently unstable. Small perturbations of an otherwise monotonic wind velocity structure tend to grow into shocks. As a result the wind of this type of star is a mixture of rarified hot material and colder dense regions, with ``cool'' and ``hot'' gas in dynamical equilibrium. It is common in the literature to attribute wind ``clumps'' to dense formations with temperatures much lower than the temperatures of the hot gas. These clumps are presumably responsible for observed line-profile variability or the formations of DACs (discrete absorption components) observed in the optical and UV spectra of O stars. The estimated quantity of these clumps is of the order $10^{3}$ (Howk et al. 2002); their size hierarchy varies and appears dependant on spectral type and possible structure of local magnetic fields. The largest DACs can be seen in the line spectrum of the star for $\sim$few days but do not exist longer than a few rotational periods of the star (Kaper et al. 1997)

The whole Cyg X-1 system, including the black hole secondary, is surrounded by the relatively dense clumpy stellar wind. Its large amplitude of variability changes the column density of the neutral gas significantly in the direction of the secondary. Two types of such variability are known: orbital, that is connected with the variable column density due to orbital movement of the secondary in the envelope of the primary component (Brocksopp et al. 1999b; Wen et al. 1999); and dips, which usually last in the X-ray spectrum for several minutes up to 8 hours (Baluci\'{n}ska-Church et al. 2000). Both types of X-ray variability are seen during the low/hard state. Dips in the soft X-ray spectrum are relatively widely spread. During the dip, the spectrum is hardened and an iron K absorption edge may be seen, showing that they are due to photoelectric absorption (Kitamoto et al. 1984). Outside the dips the column density of neutral gas in the direction of secondary is $\sim6.21\pm0.22\times 10^{21} \mbox{cm}^{-2}$ (Schulz et al. 2002) but can easily reach $\sim1\times 10^{23} \mbox{cm}^{-2}$ during dips (Baluchi\'{n}ska-Church et al. 1997). The size of the clumps that are responsible for short-lived dips is $\sim10^{9}$ cm, i.e. relatively small in size. The quantity of clumps on the line of sight depends on the geometrical length of the envelope, so dips are more frequent at superior conjunction (Wen et al. 1999; Baluchi\'{n}ska-Church et al. 2000).

\begin{acknowledgements}
The authors are grateful to the anonymous referee for helping to improve this paper.
AET and CB gratefully acknowledge receipt of financial support from the Royal Society for collaborate work with the FSU countries.

\end{acknowledgements}

\end{document}